\documentclass[twocolumn,preprintnumbers,amsmath,amssymb,jcp]{revtex4}

\usepackage{graphicx}
\usepackage{dcolumn}
\usepackage{bm}
\usepackage{color}
\usepackage{multirow}
\usepackage{longtable}
\usepackage{enumerate}
\usepackage{color}
\usepackage{paralist}
\usepackage{verbatim}
\usepackage[normalem]{ulem}

\newcommand{\sandw}[3]{\langle#1|#2|#3\rangle}

\newcommand{\bra}[1]{\langle #1|}
\newcommand{\ket}[1]{|#1\rangle}

\newcommand{\eps}[0]{\varepsilon}
\newcommand{\pphi}[0]{\varphi}
\newcommand{\rr}[0]{\vec r}
\newcommand{\rrp}[0]{\vec{r}\,'}

\newcommand{\lp}[0]{\left}
\newcommand{\rp}[0]{\right}

\newcommand{\de}[0]{\partial}
\newcommand{\KS}[0]{{K\!S}}

\newcommand{\half}[0]{\frac{1}{2}}
\newcommand{\rarr}[0]{\rightarrow}

\begin{document}

\title{Fundamental gaps with approximate density functionals: \\
the derivative discontinuity revealed from ensemble considerations}

\author{Eli Kraisler}
\affiliation{Department of Materials and Interfaces, Weizmann Institute of Science, Rehovoth 76100, Israel}

\author{Leeor Kronik}
\affiliation{Department of Materials and Interfaces, Weizmann Institute of Science, Rehovoth 76100, Israel}

\date{\today}

\begin{abstract}

The fundamental gap is a central quantity in the electronic structure of matter. Unfortunately, the fundamental gap is not generally equal to the Kohn-Sham gap of density functional theory (DFT), even in principle. The two gaps differ precisely by the derivative discontinuity, namely, an abrupt change in slope of the exchange-correlation (xc) energy as a function of electron number, expected across an integer-electron point. Popular approximate functionals are thought to be devoid of a derivative discontinuity, strongly compromising their performance for prediction of spectroscopic properties. Here we show that, in fact, \emph{all} exchange-correlation functionals possess a derivative discontinuity, which arises naturally from the application of ensemble considerations within DFT, without any empiricism. This derivative discontinuity can be expressed in closed form using only quantities obtained in the course of a standard DFT calculation of the neutral system. For small, finite systems, addition of this derivative discontinuity indeed results in a greatly improved prediction for the fundamental gap, even when based on the most simple approximate exchange-correlation density functional - the local density approximation (LDA). For solids, the same scheme is exact in principle, but when applied to LDA it results in a vanishing derivative discontinuity correction. This failure is shown to be directly related to the failure of LDA in predicting fundamental gaps from total energy differences in extended systems. 
\end{abstract}


\maketitle

\section{Introduction}

Density-functional theory (DFT)~\cite{HK'64,KS'65,PY,DG,Primer,EngelDreizler2011,Burke12,Capelle_BirdsEye,AM1} is the leading theoretical framework for studying the electronic properties of matter. It is based on mapping the interacting-electron system into the Kohn-Sham (KS) system of non-interacting electrons, which are subject to a common effective potential. DFT is a first principles approach, i.e.\ the only necessary input for the theory is the external potential, $v_{ext}(\rr)$, and the total number of electrons, $N$; no experimental data are required. In principle, the KS mapping is exact. In practice, it involves an exchange-correlation (xc) density functional, $E_{xc}[n(\rr)]$, whose exact form is unknown and is always approximated.

Present-day approximations within DFT already make it widely applicable to a variety of many-electron systems in physics, chemistry, and materials science~\cite{Martin,Hafner,Kaxiras,Cramer,ShollSteckel}. Specifically, quantities that can be derived from the total energy of the system, notably structural and vibrational properties, can often be obtained with a satisfactory accuracy of a few percent or better. 
However, extraction of quantities such as the ionization potential (IP) or the fundamental gap directly from the KS eigenvalues often results in serious discrepancies with experiment (see, e.g., Refs.~\cite{Louie_review97,Chan99,AllenTozer02,KueKronik08,Teale08,Refaely-Abramson2011,Blase2011}).

It is by now well-known~\cite{PPLB82,Levy1984,Almbladh1985,PerdewLevy97} that for the exact (but generally unknown) xc functional the highest occupied ($ho$) KS eigenvalue would equal the negative of the ionization potential of the interacting-electron system. This result is known as the ionization potential theorem. However, the KS gap, $E_g^\KS$, namely the energy difference between the lowest unoccupied ($lu$) and $ho$ KS eigenvalues, would still not equal the fundamental gap of the interacting system, $E_g$. This difference is due to a finite, spatially uniform ``jump'' in the KS potential, experienced as the electron number, $N$, crosses an integer value. This ``jump'' is called the derivative discontinuity, $\Delta$, for reasons discussed in detail below. Extensive numerical investigations have shown~\cite{Godby1987,Godby1988,Chan99,AllenTozer02,Teale08} that the value of $\Delta$ associated with the exact KS potential for various systems of interest is not at all negligible in comparison to $E_g^\KS$. 
It is commonly understood that standard semi-local xc potentials, like the local density approximation (LDA)~\cite{VWN'80,PZ'81,PW'92} or generalized-gradient approximations (GGAs)~\cite{PW86,Becke1988,LYP88,PBE'96,WuC06,Haas2011a}, do not possess a derivative discontinuity by construction. As a result, such approximations effectively ``average over'' it in the vicinity of the integer point~\cite{Perdew1983,Tozer1998}. Consequently, the ionization potential theorem is grossly disobeyed and in addition the fundamental gap is greatly underestimated.

For finite systems, a correct positioning of the $ho$ and the $lu$ KS eigenvalues is highly advantageous when describing processes like ionization, photoemission, charge transport or transfer, etc.~\cite{Kronik2012,KronikKuemmel_PES,Quek2009,Dreuw2004,Tozer2003,
Stein2009,Kaduk2012}. 
But at least the ionization potential and electron affinity, and ergo the fundamental gap, which equals the difference between the two, can be calculated based on total energy differences between neutral, cation, and anion (see, e.g., \cite{Kaduk2012,Ogut97,Kronik2002a,Moseler2003,Weissker2004,Kr'10,Argaman2013})

For periodic systems, e.g., crystalline solids, this is no longer the case. Because such systems are represented by a unit cell with periodic boundary conditions, varying the number of electrons per unit cell means adding or subtracting charge from each replica of the unit cell and therefore an infinitely large charge from the system as a whole. The ensuing divergence is usually avoided by introducing a compensating uniform background to the unit cell, keeping the overall system neutral. However, this hinders the straightforward use of total energy differences for deducing fundamental gaps, as discussed, e.g., in Refs.~\cite{Sharma2008,Chan2010}. Thus, there is a clear advantage in extracting the fundamental gap based on KS eigenvalues and other quantities arising in the calculation of the neutral system itself, without alteration of the number of electrons.

Many novel approaches have been developed within DFT for improving the accuracy of fundamental gap prediction beyond that afforded by conventional semi-local functionals, with an emphasis on applications to crystalline solids. These can be broadly divided into several categories.

Within the KS framework, significant attention has been devoted to the employment of functionals that do possess an inherent derivative discontinuity. Much of the effort involved the exact-exchange (EXX) functional, using the optimized effective potential (OEP)~\cite{Grabo_MolPhys,EngelDreizler2011,KueKronik08} approach~\cite{Bylander1996a,Stadele1997,Stadele1999,Magyar2004,Gruning2006,Gruning2006a,Rinke2005,Rinke2008}. More recently, novel semi-local functionals were constructed so as to mimic EXX-OEP properties~\cite{Becke2006, Tran2007, Tran2009,Kuisma2010,Armiento2013,Laref2013}.

Alternatively, it is possible to step outside the KS framework and use the generalized KS (GKS) scheme~\cite{Seidl1996,KueKronik08,Baer2010,Kronik2012}, where mapping to a partially interacting electron system results in a KS-like equation that includes a non-multiplicative potential operator. This operator can reduce the magnitude of the derivative discontinuity, potentially driving it down to negligible values. Practical GKS schemes often rely on the Fock operator, or variants thereof. Such GKS calculations were performed, e.g.,  
by employing the screened exchange approach~\cite{Seidl1996,Picozzi2000,Geller2001,Stampfl2001,Robertson2006,Lee2006},
by using global hybrid functionals~\cite{Dovesi2000,Bredow2000,Muscat2001,Cora2004,Paier2006b,Paier2006,
Moses2010,Moses2011,Sai2011,Jain2011,Moussa2012,Sai2011},
range-separated hybrid functionals~\cite{Heyd2005,Krukau2006,Paier2006b,Paier2006,Gerber2007,Eisenberg2009,Clark2011, Henderson2011,Schimka2011a,Lucero2012,Stein2010,Refaely-Abramson2013}, and by applying a scaling correction to the Hartree and exchange functionals~\cite{Zheng11,Zheng2013}. Alternatively, one can step outside the KS scheme by introducing orbital-specific corrections, where different electrons of the KS system are subject to different potentials. This is achieved, e.g., using self-interaction correction methods~\cite{PZ'81,Svane1990,Heaton1983,Filippetti2003}, DFT+U and Koopmans' compliant functionals~\cite{Anisimov1997,Cococcioni2005, Janotti2006, Lany2008, Lany2009, Forti2012, Andriotis2013,Dabo2010,Dabo2013a,Dabo2013}, the LDA-1/2 method~\cite{Ferreira2008}, Fritsche's generalized approach~\cite{Fritsche1991,Remediakis1999}, or a scissors-like operator to the KS system that affects only the vacant states~\cite{Mera2009}.

A different possibility altogether is to sidestep the full charging problem by considering total energy differences arising from appropriately constructed partial charging schemes, e.g., by employing dielectric screening properties~\cite{Chan2010}, by averaging the KS transition energies around the direct band-gap transition~\cite{Scharoch2013}, or by employing perturbative curvature considerations~\cite{Stein2012a}.

But must conventional semi-local functionals really be abandoned as far as band gap prediction is concerned? Recently, this question has received some attention. For finite systems, Andrade and Aspuru-Guzik \cite{Andrade2011a} and Gidopoulos and Lathiotakis~\cite{Gidopoulos2012} have suggested derivative-discontinuity correction schemes based on an electrostatic correction of the asymptotic potential~\cite{Leeuwen1994a}. Chai and Chen~\cite{Chai2013} derived a perturbative approach for the evaluation of the missing derivative discontinuity. In the first order, this perturbative treatment leads to the "frozen orbital approximation" result, discussed lately by Baerends and co-workers~\cite{Baerends2013}. We have shown that, contrary to conventional wisdom, in fact the KS potential derived from \emph{any} xc functional possesses a derivative discontinuity~\cite{KrKronik13}, whose value emerges naturally and non-empirically from the ensemble generalization of DFT~\cite{Lieb,RvL_adv,PPLB82,Joubert2013,Hellgren2012,Hellgren2013}. These approaches have, so far, been applied to atoms and molecules and their potential for the solid-state remains unexplored.

Here, we derive an explicit, closed-form expression for the derivative discontinuity, $\Delta$, of an arbitrary many-electron system studied with an arbitrary xc functional. This derivation is based on the ensemble generalization of the Hartree and xc energy terms suggested in Ref.~\cite{KrKronik13}. Furthermore, $\Delta$ is expressed using only quantities associated with the neutral system, thereby avoiding alteration of the electron number. Therefore, the formalism is, in principle, applicable to both finite and periodic systems. Focusing on the latter, we explore analytically the scaling of $\Delta$ with system size. We find that while for the exact xc functional $\Delta$ must be independent of system size, for standard xc approximations like the LDA the derivative discontinuity vanishes. This failure is shown to directly related to the failure of LDA in predicting fundamental gaps from total energy differences in extended systems. These findings are demonstrated by illustrative calculations.

\section{The ensemble approach} \label{sec.ensemble}

The central quantity we discuss below is the fundamental gap, $E_g$. It is defined as
\begin{equation}\label{eq.G.def}
E_g = I - A,
\end{equation}
i.e., it is the difference between the ionization energy, $I$, and the electron affinity, $A$. As $I$ and $A$ involve removal and addition of an electron, respectively, in the following we analyse in detail the properties of a many-electron system with a varying number of electrons.

\begin{figure}
  \includegraphics[scale=0.50, trim=15mm 0mm 0mm 0mm]{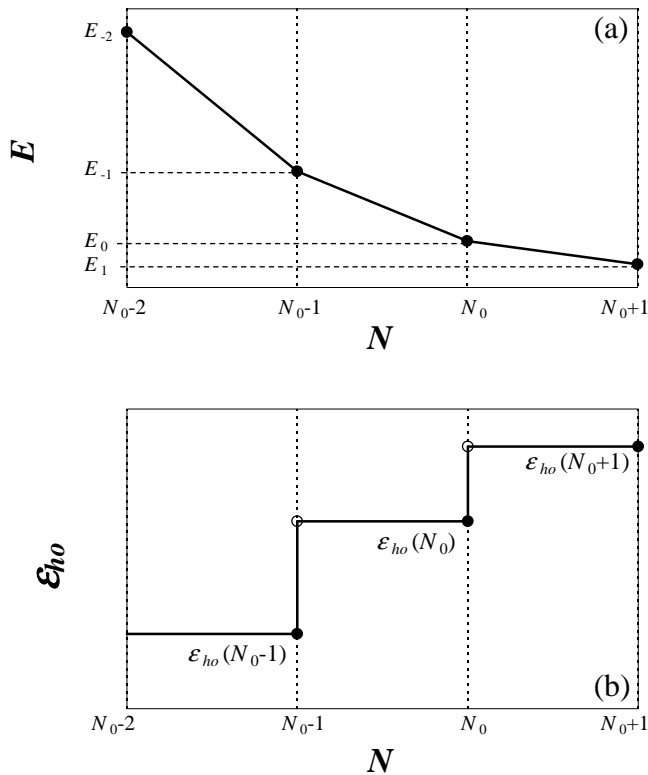}\\
  \caption{A schematic plot of the total energy $E$ (top) and the highest occupied KS energy level $\eps_{ho}$ (bottom) depending on the number of particles $N$ for the exact xc-functional}\label{fig.piecewise}
\end{figure}

At zero temperature, the ground state of a many-electron system with a (possibly) fractional number of electrons $N = N_0 + \alpha$ ($N_0 \in \mathbb{N}$ and $\alpha \in [0,1]$) is described by an ensemble state~\cite{PPLB82} 
\begin{equation}\label{eq.Lambda}
\hat{\Lambda} = (1-\alpha)\ket{\Psi_{N_0}}\bra{\Psi_{N_0}} + \alpha\ket{\Psi_{N_0+1}}\bra{\Psi_{N_0+1}},
\end{equation}
where $\ket{\Psi_{N_0+p}}$ is a pure many-electron ground state with $N_0+p$ electrons and $p$ is 0 or 1
~\footnote{Here and below it is assumed that the ground states of the system of interest, of its anion and of its cation are not degenerate, or that the degeneracy can be lifted by applying an infinitesimal external field},~\footnote{The fact that Eq.~(\ref{eq.Lambda}) includes contributions only from the $N_0$- and the $N_0+1$-states relies on the conjecture that the series $E(N_0)$ for $N_0 \in \mathbb{N}$ is a convex, monotonously decreasing series. In other words, all ionization energies $I(N_0):=E(N_0-1)-E(N_0)$ are positive, and higher ionizations are always larger than the lower ones: $I(N_0-1) > I(N_0)$. This conjecture, although strongly supported by experimental data, remains without proof, to the best of our knowledge~\cite{DG,Lieb,Cohen12}.}. 
As a result, the ground-state energy $E(N)$ at a fractional $N$ is a linear combination of the ground-state energies at the closest integer points:
\begin{equation}
E(N) = (1-\alpha)E(N_0) + \alpha E(N_0+1).
\end{equation}
Therefore, the function $E(N)$ is piecewise-linear (see Fig.~\ref{fig.piecewise}(a) for an illustration).

The energy obtained in DFT using the KS system with the exact xc functional has to reproduce the piecewise-linear behavior. Janak's theorem~\cite{Janak1978} states that the $i$-th KS eigenenergy, $\eps_i$, equals $\de E / \de f_i$ -- the derivative of the total energy of the interacting system, $E$, with respect to the occupation of the $i$-th level, $f_i$. Applying this theorem, we find that with the exact xc functional the $ho$ eigenenergy, $\eps_{ho}$, is a stair-step function of the number of electrons, $N$ (see Fig.~\ref{fig.piecewise}(b)).

A non-vanishing fundamental gap in a many-electron system indicates a discontinuity in the chemical potential, namely, the cost of electron insertion and removal is different. This physical discontinuity is manifested as a mathematical discontinuity exhibited in Fig.~\ref{fig.piecewise}. From the perspective of total energies, $I= E(N_0-1) - E(N_0)$ and $A = E(N_0) - E(N_0+1)$ are the slopes of $E(N)$ to the left and right of $N_0$, respectively, which are generally different from each other. The gap $E_g$ is then the difference in these two slopes. From the perspective of KS eigenvalues, $I = - \eps_{ho}(N_0)$ and $A = - \eps_{ho}(N_0+1)$ (owing to the IP-theorem), and then $E_g$ is the magnitude of the step in $\eps_{ho}(N)$ at $N_0$~\cite{PPLB82,Levy1984,Almbladh1985,PerdewLevy97}. In other words, $E_g$ is the discontinuity of the {\it derivative} of the $E(N)$ curve, or the discontinuity of the $\eps_{ho}(N)$ curve itself, at $N_0$. 

Clearly, in order to obtain the fundamental gap from total energy differences one has to calculate not only the system of interest (with $N=N_0$), but also its anion ($N=N_0+1$) and its cation ($N=N_0-1$). Figure~\ref{fig.piecewise}(b) suggests, however, an alternative route: $E_g$ can, at least in principle, be derived by analyzing the left and the right limits, $N \rarr N_0^-$ and $N \rarr N_0^+$, of the neutral many-electron system.

Consider now the celebrated KS equation~\cite{KS'65}, given in Hartree units by 
\begin{equation}
\lp( -\frac{1}{2} \nabla^2 + v_\KS(\rr) \rp) \pphi_i(\rr) = \eps_i \pphi_i(\rr),
\end{equation}
where $v_\KS(\rr)$ is the KS potential and $\pphi_i(\rr)$ are the KS orbitals. 
What elements of the KS equation may cause the ``jump'' of $\eps_{ho}(N)$ at integer $N$, shown in Fig.~\ref{fig.piecewise}(b)? One obvious source is simply the fact that the state named $ho$ for $N \rarr N_0^+$ is \emph{not} the same state as the one named $ho$ for $N \rarr N_0^-$, due to the infinitesimal occupation of the next available energy level, whose eigenvalue is different.
However, there is a more subtle second source: There is nothing to prevent the KS potential itself from exhibiting an abrupt ``jump'' across the integer point~\cite{PPLB82,Perdew1983,Sham1983,KueKronik08}. Because an infinitesimal change in $N$ across the integer point can only change the density $n(\rr)$ infinitesimally~\cite{Perdew1983}, the potential $v_\KS(\rr)$ cannot ``jump'' by more than a spatial constant. Otherwise, in the limit of $N_0$ the same density would be achieved from two potentials differing by more than a constant, in direct contradiction of the Hohenberg-Kohn theorem~\cite{HK'64}. In the complementary energy picture, the first source mentioned above~-- change in leading orbital~-- results in an abrupt slope change of the KS kinetic energy, $T_\KS = \sum_i f_i \sandw{\pphi_i}{-\half \nabla^2}{\pphi_i}$. The second source~-- the ``jump'' in the KS potential~-- creates an abrupt slope change in the Hartree-exchange-correlation energy, $E_{Hxc}$.

We denote the aforementioned spatial constant by $\Delta$ and write: 
\begin{equation}\label{eq.vR.vL}
v_\KS ^R (\rr) = v_\KS ^L (\rr) + \Delta.
\end{equation}
Here and below we use the superscripts $^L$ and $^R$ to denote quantities \emph{immediately} to the left or to the right of the integer point $N_0$. 

Equation~(\ref{eq.vR.vL}) has two immediate consequences. Upon infinitesimal crossing of the integer point, $N_0$, from $N_0^-$ to $N_0^+$,
all the KS eigenvalues ``jump'' by the same quantity, i.e., $\eps_i ^R = \eps_i ^L + \Delta$. However, the KS orbitals do not exhibit any change: $\pphi_i^R(\rr) = \pphi_i^L(\rr)$. As a special case of these statements, $\pphi_{ho}^R(\rr) = \pphi_{lu}^L(\rr)$ and $\eps_{ho} ^R = \eps_{lu} ^L + \Delta$.
These simple statements are key to the following derivation. For the gap we then obtain
\begin{equation}\label{eq.G.def.with.Delta}
E_g = \eps_{ho}(N_0+1) - \eps_{ho}(N_0) = \eps_{ho}^R - \eps_{ho}^L = \eps_{lu}^L - \eps_{ho}^L + \Delta.
\end{equation}
Using the definition of the KS gap, $E_g^\KS = \eps_{lu}^L - \eps_{ho}^L$, we arrive at~\cite{Perdew1983,Sham1983,PerdewLevy97,Teale08}
\begin{equation}\label{eq.G.def.with.Delta2}
E_g = E_g^\KS + \Delta.
\end{equation}
The above expression is an exact result and therefore must be obeyed by results obtained from the exact KS potential. For any given approximate xc potential, however, the degree to which Eq.~(\ref{eq.G.def.with.Delta}) is obeyed may vary, depending on the deviation of $\eps_{ho}(N)$ from flatness, or equivalently, on the deviation of $E(N)$ from piecewise-linearity~\cite{MoriS06,Stein2012a,Ruzsinszky07,Vydrov07,Cohen08,HaunScu10,Cohen12,Srebro2012a,Gledhill2013,
Hofmann2013}.

As already mentioned, the density is continuous across the integer point, i.e., $n^R(\rr) = n^L(\rr)$. Because in conventional (semi-)local approximate xc functionals such as the LDA and GGAs, the xc potential is a continuous function of the density (and its gradient), it is commonly believed that there is no mathematical possibility for the KS potential to ``jump'' and therefore $\Delta = 0$. It is this last statement that we challenge in this work. 

If the interacting-electron system has a fractional electron number, its corresponding KS system must 
also have a fractional electron number. Therefore, not only the ground state of the interacting system, but also the ground state of the KS system must unavoidably be described in terms of an ensemble, while still being fully described by a single KS potential. In analogy to Eq.~(\ref{eq.Lambda}), the KS ensemble state must be written in the form
\begin{equation}\label{eq.LambdaKS}
\hat{\Lambda}^\KS = (1-\alpha)\ket{\Phi_{N_0}^{(\alpha)}}\bra{\Phi_{N_0}^{(\alpha)}} + \alpha\ket{\Phi_{N_0+1}^{(\alpha)}}\bra{\Phi_{N_0+1}^{(\alpha)}},
\end{equation}
where $\ket{\Phi_{N_0}^{(\alpha)}}$ and $\ket{\Phi_{N_0+1}^{(\alpha)}}$ are pure KS ground states, with $N_0$ and $N_0+1$ electrons, respectively. These pure states are Slater determinants formed from the $N_0$ or $N_0+1$ occupied KS orbitals, obtained from the {\it same} KS potential, i.e., the two Slater determinants differ {\it only} in that the $N_0+1$ one contains one more orbital. The KS potential generating them is, generally, neither that of the pure $N_0$ system nor that of the pure $N_0+1$ system.  Therefore, in contrast to the quantities $\ket{\Psi_{N_0+p}}$ used 
to describe the ensemble state of the interacting system, all quantities of the KS ensemble in Eq.~(\ref{eq.Lambda}) may generally change with the electron fraction, i.e., are $\alpha$-dependent. We emphasize this by using the superscript $(\alpha)$.
Ensemble-averaging the many-electron Coulomb operator $\hat W = \half \sum_i \sum_{j \neq i} | \rr_i - \rr_j|^{-1}$ in the KS system, the Hartree-exchange-correlation energy functional generalizes to ensembles in the following form~\cite{KrKronik13}:
\begin{equation}\label{eq.ET.gen}
    E_{e-Hxc}[n] = (1-\alpha) E_{Hxc}[\rho_0^{(\alpha)}] + \alpha E_{Hxc}[\rho_1^{(\alpha)}].
\end{equation}
Here, the index $e-$ signifies that the functional is ensemble-generalized, $E_{Hxc}[\rho]$ 
is the pure-state Hartree-exchange-correlation energy functional, and $\rho_p^{(\alpha)}(\rr)$ is defined as $\rho_p^{(\alpha)}(\rr) = \sum_{i=1}^{N_0+p} |\pphi_i^{(\alpha)}(\rr)|^2$, namely the sum of the squares of the first $N_0+p$ KS orbitals. We stress that $\rho_p^{(\alpha)}(\rr)$ are auxiliary quantities that are not associated with any physical density, except when $N$ is an integer. We further emphasize that the ensemble-generalized form of Eq.\ (\ref{eq.ET.gen}) is not an ansatz, but rather an inevitable consequence of employing the ensemble approach to describe a KS system of fractional number of particles. If the exact pure-state xc functional $E_{xc}[n]$ were to be inserted into Eq.~(\ref{eq.ET.gen}), the ensemble-generalized total energy would have been \emph{exactly} piecewise linear. Even then the auxiliary densities $\rho_0^{(\alpha)}$ and $\rho_1^{(\alpha)}$ need not equal the pure-state densities of $N_0$ - and $N_0+1$-systems).

The density $n(\rr)$ of the ensemble state is expressed in terms of $\rho_p^{(\alpha)}(\rr)$ as
\begin{equation}\label{eq.n.ens}
n(\rr) = (1-\alpha) \rho_0^{(\alpha)}(\rr) + \alpha \rho_1^{(\alpha)}(\rr).
\end{equation}
To obtain $E_{e-Hxc}[n]$, we construct $\rho_0^{(\alpha)}(\rr)$ and $\rho_1^{(\alpha)}(\rr)$ from the KS orbitals as mentioned above, substitute them into the functional $E_{Hxc}$ to obtain $E_{Hxc}[\rho_0^{(\alpha)}]$ and $E_{Hxc}[\rho_1^{(\alpha)}]$, and take the linear combination of the latter according to Eq.~(\ref{eq.ET.gen}). Note that this procedure is \emph{not} equivalent to constructing the ensemble density $n(\rr)$ from a linear combination of $\rho_0^{(\alpha)}(\rr)$ and $\rho_1^{(\alpha)}(\rr)$ (cf.\ Eq.~(\ref{eq.n.ens})) and substituting it into $E_{Hxc}$, as the latter functional is not linear with respect to the density.

The generalization in Eq.~(\ref{eq.ET.gen}) is applicable to \emph{any} xc functional and makes the Hartree and the xc energy components \emph{explicitly} dependent on the KS orbitals and on $\alpha$. However, there may still remain an \emph{implicit} non-linear dependence of $E_{e-Hxc}[n]$ on $\alpha$ because the KS orbitals, $\pphi_i^{(\alpha)}(\rr)$, may themselves change with $\alpha$. Finally, note that for pure states, i.e., for $\alpha=0$ or $1$, the ensemble generalized $E_{e-Hxc}[n]$ reduces to the pure-state form $E_{Hxc}[n]$, as expected.

Because the KS potential, and specifically its behavior around an integer electron number, is central to this work, we address it here in detail. Due to the ensemble generalization of the Hartree-exchange-correlation energy functional, the KS potential is expressed as
$v_\KS (\rr) = v_{ext}(\rr) + v_{e-Hxc}[n](\rr)$, where $v_{ext}(\rr)$ is the external potential and $v_{e-Hxc}[n](\rr):=\delta E_{e-Hxc} / \delta n$ is the ensemble-generalized Hartree-exchange-correlation potential. While deriving the latter from Eq.~(\ref{eq.ET.gen}) we emphasize an unusual property of $E_{e-Hxc}[n]$: it explicitly depends on $\alpha$. Therefore, the ensemble-generalized Hartree-exchange-correlation potential reads
\begin{equation}\label{eq.v_eHxc}
    v_{e-Hxc}[n](\rr) = \lp( \frac{\de E_{e-Hxc}}{\de \alpha} \rp)_n \frac{\delta \alpha}{\delta n} + \lp( \frac{\delta E_{e-Hxc}}{\delta n(\rr)} \rp)_\alpha.
\end{equation}
Since $\alpha[n] = N - {\rm floor}(N)$ and $N=\int n d^3r$, we find $\delta \alpha / \delta n = 1$. Therefore, $v_{e-Hxc}[n](\rr)$ is a sum of two terms: $v_0[n]=\lp( \de E_{e-Hxc} / \de \alpha \rp)_n$ and $v_1[n](\rr)=\lp( \delta E_{e-Hxc} / \delta n(\rr) \rp)_\alpha$. 
Because for fractional $N$ the functional $E_{e-Hxc}$ is orbital-dependent, via the quantities $\rho_p^{(\alpha)}(\rr)$, irrespective of the underlying xc functional, the potential $v_1[n](\rr)$ has to be treated with the OEP approach~\cite{Grabo_MolPhys,EngelDreizler2011,KueKronik08}. The somewhat unusual term $v_0$ is spatially uniform but $\alpha$-dependent, and it arises from the aforementioned explicit dependence of $E_{e-Hxc}[n]$ on $\alpha$. 

We focus now on $v_0$, which can be written as
\begin{align}\label{eq.v0}
\nonumber    v_0 = \lp( \frac{\de E_{e-Hxc}}{\de \alpha} \rp)_n &= \lp( \frac{\de E_{e-Hxc}}{\de \alpha} \rp)_{ \{ \pphi_i \} } \\
                 - & \int d^3r \lp( \frac{\delta E_{e-Hxc}}{\delta n(\rr)} \rp)_\alpha \lp( \frac{\de n(\rr)}{\de \alpha} \rp)_{ \{ \pphi_i \} }.
\end{align}
This result is obtained by taking the partial derivative $ ( \de E_{e-Hxc} / \de \alpha )_{ \{ \pphi_i \} }$, followed by isolation of $v_0$.
Using Eqs.~(\ref{eq.ET.gen}) and~(\ref{eq.n.ens}) to evaluate the first and the second terms on the right-hand side of Eq.~(\ref{eq.v0}), respectively, we obtain
\begin{equation}\label{eq.v0.2}
    v_0 = E_{Hxc}[\rho_1^{(\alpha)}] - E_{Hxc}[\rho_0^{(\alpha)}] - \int d^3r |\pphi_{ho}(\rr)|^2 v_1[n](\rr).
\end{equation}
for $N \in (N_0,N_0+1]$. For $N \in (N_0-1,N_0]$, one has to substitute $\rho_0^{(\alpha)}$ with $\rho_{-1}^{(\alpha)}$ and $\rho_1^{(\alpha)}$ with $\rho_0^{(\alpha)}$. 
We stress that $v_0$ is a well-defined, rather than an arbitrary, potential shift. It must be taken into account for the ensemble-generalized functional, if $\eps_{ho}$ is to equal $\de E/ \de N$, i.e., if Janak's theorem~\cite{Janak1978} is to be obeyed. 
The existence of a spatially uniform potential shift $v_0$ is in agreement with earlier studies~\cite{Perdew1983,Parr1983}, which found that whereas for fractional $N$ the KS potential is well-defined, for integer $N$ it is defined up to a constant. The latter ambiguity in the definition of the KS potential can be removed by reaching the integer number of electrons $N_0$ from below (for a discussion, see~\cite{Levy1984,PerdewLevy97}).

Note that $v_0$ and $v_1(\rr)$ are obtained via different quantities when $N \in (N_0-1,N_0]$ and $N \in (N_0,N_0+1]$. Therefore, when approaching $N_0$ from the left and from the right, we generally expect to obtain different KS potentials. In other words, we expect $v_\KS(\rr)$ to \emph{change discontinuously when crossing an integer number of electrons}. As mentioned above, this discontinuity must be a spatially uniform constant, $\Delta$ (cf.\ Eq.~(\ref{eq.vR.vL})).

\begin{figure}
 \includegraphics[scale=0.50, trim=10mm 0mm 0mm 0mm]{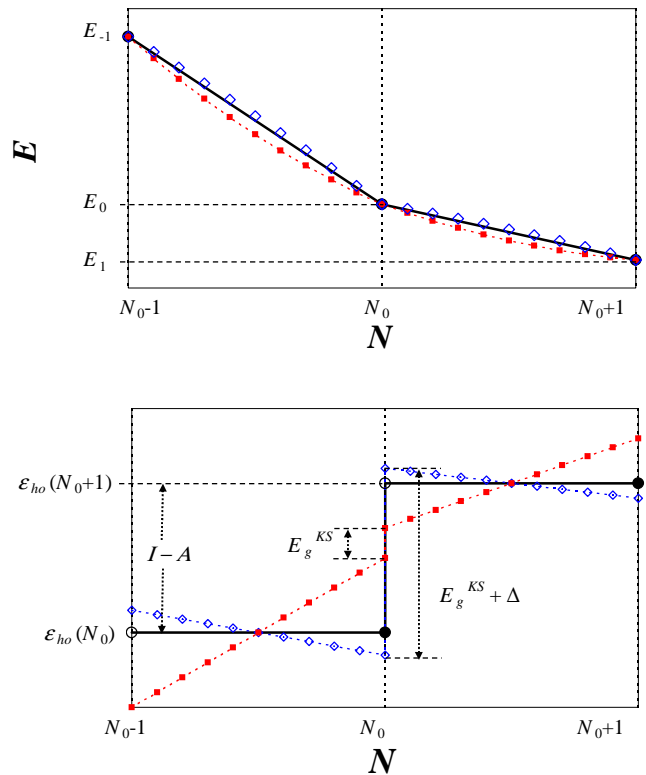}\\
\caption{A schematic plot of the total energy $E$ (top) and the highest occupied KS energy level $\eps_{ho}$ (bottom) as a function of the number of electrons, $N$, obtained with the exact functional (solid line), with an approximate functional (e.g. the LDA; squares) and with the ensemble generalization of the approximate functional (rhombi). The KS gap, $E_g^\KS$, the ensemble-corrected gap, $E_g^\KS + \Delta$ (Eq.~(\ref{eq.G.def.with.Delta2})), and the exact gap, $I-A$ (Eq.~(\ref{eq.G.def})), are denoted on the figure.}\label{fig.piecewise.ensemble}
\end{figure}

The consequences of the generalization presented above 
are schematically depicted in Fig.~\ref{fig.piecewise.ensemble}, based on numerical results for finite systems presented in Ref.~\cite{KrKronik13}. The ensemble generalization brings $\eps_{ho}(N)$ closer to the desired stair-step form: it becomes more flat for fractional $N$ and the ``jump'' experienced at integer $N$ is increased by $\Delta$. As observed by Stein~\emph{et~al.}~\cite{Stein2012a}, the derivative discontinuity and piecewise-linearity of the total energy are two sides of the same coin: A missing derivative discontinuity must be accompanied by deviation from piecewise linearity, and vice versa. Therefore, improvement in the description of $\eps_{ho}(N)$ inevitably reflects on the total energy curve: the spurious convexity of $E(N)$ is significantly reduced, bringing it closer to the desired piecewise-linear behavior, and the abrupt change of slope near the integer points is better reproduced. 
Importantly, the numerical results given in Ref.~\cite{KrKronik13} show that for ions of atoms and small molecules ensemble-generalization of the local spin-density approximation (LSDA) does indeed yield fundamental gaps in much better agreement with experiment than standard LSDA calculations. For example, for H$_2^+$, a gap of 5.80 eV predicted with LSDA is increased to 17.96 eV with ensemble-LSDA, reducing the discrepancy with respect to experiment from 70\% to 8\%. For C$^+$, LSDA predicts a gap of 0.26 eV, which is increased to 15.31 eV with ensemble-LSDA, reducing the discrepancy with experiment from 98\% to 17\%.

Nonetheless, for an approximate xc functional the ensemble-corrected gap $E_g = E_g^\KS + \Delta$, still does not exactly equal $I-A$. The difference that remains is due to a deviation of $\eps_{ho}(N)$ from flatness, which is attributed to the implicit non-linear dependence of an approximate $E_{e-Hxc}[n]$ on $\alpha$.

\section{The derivative discontinuity $\Delta$} \label{sec.dd}

In this section, an analytical expression for the derivative discontinuity $\Delta$ is derived by taking the limits $N \rarr N_0^-$ and $N \rarr N_0^+$. First, let us introduce some notation. When $N \rarr N_0^+$, i.e.\ $\alpha \rarr 0^+$ (the $^R$ limit), the quantity $\rho_0^{(0)}$ is termed $n_0$, and the quantity $\rho_1^{(0)}$ is termed $n_1$. 
Note that because $\pphi_i^R(\rr) = \pphi_i^L(\rr)$, these quantities are continuous when crossing $N_0$; for this reason they do not receive the index~$^R$.
Also note that the energy $E_{e-Hxc}[n]$ and the density $n$ were defined in Eqs.~(\ref{eq.ET.gen}) and~(\ref{eq.n.ens}) for the case when $N \in [N_0, N_0+1]$, i.e.\ to the right of the point $N_0$. In the region $[N_0-1, N_0]$, which is of interest here as well, these quantities are defined similarly, substituting $\rho_0^{(\alpha)}$ with $\rho_{-1}^{(\alpha)}$ and $\rho_1^{(\alpha)}$ with $\rho_0^{(\alpha)}$. Furthermore, when $N \rarr N_0^-$, i.e.\ $\alpha \rarr 1^-$ (the $^L$ limit), the quantity $\rho_{-1}^{(1)}$ is denoted $n_{-1}$.

We now focus on $\Delta = v_{e-Hxc}^R(\rr) - v_{e-Hxc}^L(\rr)$, which we choose to express as $\Delta = \Delta_0 + \Delta_1$. Here
\begin{equation}\label{eq.D0}
\Delta_0 = v_0^R - v_0^L
\end{equation}
 and
\begin{equation}\label{eq.D1}
\Delta_1 = v_1^R(\rr) - v_1^L(\rr).
\end{equation}
Reaching the point $N=N_0$ from the left, we obtain from Eq.~(\ref{eq.v0.2})
\begin{equation}\label{eq.v0L}
    v_0^L = E_{Hxc}[n_0] - E_{Hxc}[n_{-1}] - \int d^3r |\pphi_{ho}^L(\rr)|^2 v_1^L[n_0](\rr).
\end{equation}
Reaching $N_0$ from the right, we obtain similarly
\begin{equation}\label{eq.v0R-prelim1}
    v_0^R = E_{Hxc}[n_1] - E_{Hxc}[n_0] - \int d^3r \,\, |\pphi_{ho}^R(\rr)|^2 v_1^R[n_0](\rr).
\end{equation}
Recalling that $\pphi_{ho}^R(\rr) = \pphi_{lu}^L(\rr)$ and using Eqs.~(\ref{eq.D1}) and~(\ref{eq.v0R-prelim1}), we rewrite $v_0^R$ as
\begin{equation}\label{eq.v0R}
    v_0^R = E_{Hxc}[n_1] - E_{Hxc}[n_0] - \Delta_1 - \int d^3r \,\, |\pphi_{lu}^L(\rr)|^2 v_1^L[n_0](\rr).
\end{equation}
When reaching $N_0$ from the left, $v_1^L[n_0](\rr) = v_{Hxc}[n_0](\rr)$, where $v_{Hxc} = \delta E_{Hxc}/ \delta n$ is the usual Hartree-exchange-correlation potential defined for the pure ground state with $N_0$ electrons.

Finally, subtracting Eq.~(\ref{eq.v0L}) from Eq.~(\ref{eq.v0R}) and using Eqs.~(\ref{eq.D0}) and~(\ref{eq.D1}), we can express the discontinuity $\Delta$ solely in terms of the $^L$ quantities, i.e.\ using only quantities that correspond to the system of interest, with exactly $N_0$ electrons. $\Delta$ is given below, suppressing now the index $^L$ for clarity:
\begin{align}\label{eq.Delta}
\nonumber    \Delta = E_{Hxc}[n_1] &- 2E_{Hxc}[n_0] + E_{Hxc}[n_{-1}] \\
    &+ \int  d^3r \,\, v_{Hxc}[n_0] \lp( |\pphi_{ho}(\rr)|^2 - |\pphi_{lu}(\rr)|^2 \rp)
\end{align}
Equation~(\ref{eq.Delta}) is a key result of the current contribution. It is achieved completely from first principles, meaning that no approximations were introduced during its derivation. 
Because the derivation is valid for an arbitrary xc functional (exact or approximate), we conclude that \emph{all} xc functionals possess a generally non-zero derivative discontinuity $\Delta$, which is revealed by rigorous employment of the ensemble approach in DFT. This includes, of course, also the simplest xc approximation -- the LDA, used for the computations of Ref.\ \cite{KrKronik13} and in calculations presented below.

Importantly, $\Delta$ as expressed in Eq.~(\ref{eq.Delta}) is obtained using only quantities that belong to the original system of interest with $N_0$ electrons. Therefore its calculation does not require any alteration of the number of electrons in the system. In particular, it is also applicable to periodic systems, namely, ``jellium'' background charge corrections do not have to be considered in Eq.\ \ref{eq.Delta} because it is derived from a limit around the equilibrium point rather than from actual addition of charges.

It is well-known (see, e.g.,~\cite{ChelikCohen92}) that although the band structure of the KS system cannot be rigorously related to properties of the interacting system, it nevertheless can serve as an approximation to the charged excitation spectrum of the latter, apart from a rigid shift of the unoccupied bands with respect to the occupied ones. The corresponding shift is usually introduced empirically, or by relying on theories beyond DFT, e.g.\ many-body perturbation theory, and bears the name of the "scissors shift". Here, $\Delta$ provides a similar effect, with the important difference that it is derived  completely within DFT.

The derivation above was performed within the OEP framework. However, it is important to note that the calculation of $\Delta$ does \emph{not} require any actual use of the OEP formalism, but requires only simple operations of negligible numerical effort with quantities readily available from a routine DFT calculation. Actual employment of the OEP scheme is needed only for calculation of the $E(N)$ curve for fractional $N$.

\section{The limit of an infinitely large system}\label{sec.solid}

As discussed in the preceding section, Eq.~(\ref{eq.Delta}) is applicable in principle to both finite and infinite systems. In this section, we investigate the properties of $\Delta$ for a periodic system by considering how it scales with system size as the latter approaches infinity. We obtain the analytical limiting expression and address its properties for both the exact exchange-correlation potential and the local density approximation (LDA).

Consider a many-electron system, whose external potential, $v_{ext}(\rr)$, is periodic in space, i.e., $v_{ext}(\rr + \vec{R}) = v_{ext}(\rr)$, where $\vec{R}$ is a Bravais lattice vector. Neglecting surface effects, all properties of this system, including its derivative discontinuity $\Delta$, can be obtained from the limit of a collection of $M$ unit cells as $M \rarr \infty$. Let us define some terms that are useful for taking such a limit. The total number of electrons the system is $M N_0$, where $N_0$ is the (finite) number of electrons per unit cell. The electron density is $n_0(\rr) = \sum_{i=1}^{M N_0} |\pphi_i(\rr)|^2$. The KS orbitals $\pphi_i(\rr)$ are, as usual, normalized to 1 when integrating over the whole system, i.e., $\int_{all} |\pphi_i(\rr)|^2 d^3r =1$, where the subscript $all$ denotes integration over the entire system. Therefore, $\int_{all} n_0(\rr) d^3r = M N_0$ as appropriate. Integration over one unit cell yields $\int_{u.c.} n_0(\rr) d^3r = N_0$ and $\int_{u.c.} |\pphi_i(\rr)|^2 d^3r =M^{-1} \rarr 0$, where the subscript $u.c.$ denotes integration over one unit cell. We therefore define a renormalized KS orbital, $\bar{\pphi}_i(\rr) = \sqrt{M} \pphi_i(\rr)$, such that $\int_{u.c.} |\bar{\pphi}_i(\rr)|^2 d^3r = 1$.  Like the electron density, $| \bar{\pphi}_i(\rr)| ^2 $ remains finite for large $M$. 

To assess the limiting form of Eq.~(\ref{eq.Delta}), we first address
$E_{Hxc}^{all}[n_0+|\pphi_{lu}|^2]$, which can be written as $E_{Hxc}^{all}[n_0+ \frac{1}{M}|\bar{\pphi}_{lu}|^2]$ using the renormalized orbitals. The Hartree-exchange-correlation energy can then be Taylor-expanded around $n_0$, with $1/M$ serving as the small parameter, in the form:
\begin{align}
\nonumber 	& E_{Hxc}^{all} \lp[ n_0 + \frac{1}{M}|\bar{\pphi}_{lu}|^2 \rp] = E_{Hxc}^{all}[n_0] + \frac{1}{M} \int_{all} \!\! d^3r \left. \frac{\delta E_{Hxc}}{\delta n(\rr)} \rp|_{n_0} \cdot \\
\nonumber   & |\bar{\pphi}_{lu}(\rr)|^2 + \frac{1}{2M^2} \int_{all} \!\! d^3r \int_{all} \!\! d^3r' \left. \frac{\delta^2 E_{Hxc}}{\delta n(\rr) \delta n(\rrp)} \rp|_{n_0} \cdot \\
			& |\bar{\pphi}_{lu}(\rr)|^2 |\bar{\pphi}_{lu}(\rrp)|^2  + O \lp( \frac{1}{M^3} \rp)
\end{align}
A similar expression can be easily written for $E_{Hxc}^{all}[n_0-|\pphi_{ho}|^2]$. Denoting the Hartree-exchange-correlation kernel by $f_{Hxc}[n](\rr,\rrp) := \delta^2 E_{Hxc} / \delta n(\rr) \delta n(\rrp)$, recognizing that $\delta E_{Hxc} / \delta n(\rr) = v_{Hxc}[n](\rr)$, and using the renormalized orbitals $\bar{\pphi}_i(\rr)$ in Eq.~(\ref{eq.Delta}), we obtain: 
\begin{align}\label{eq.Delta.M}
\nonumber & \Delta = \frac{1}{2M^2} \int_{all} \!\!\!\! d^3r \int_{all} \!\!\!\! d^3r'  f_{Hxc}[n_0](\rr,\rrp) [ |\bar{\pphi}_{lu}(\rr)|^2  
|\bar{\pphi}_{lu}(\rrp)|^2 \\ 
& +  |\bar{\pphi}_{ho}(\rr)|^2 |\bar{\pphi}_{ho}(\rrp)|^2 ] + O \lp( \frac{1}{M^3} \rp)
\end{align}

The Hartree-exchange-correlation kernel $f_{Hxc}[n](\rr,\rrp)$ can be written as a sum of the Hartree and xc components: $f_{Hxc}[n](\rr,\rrp) = f_{H}[n](\rr,\rrp) + f_{xc}[n](\rr,\rrp)$, where $f_{H}[n_0](\rr,\rrp) = \delta^2 E_{H} / \delta n(\rr) \delta n(\rrp) = 1 / |\rr - \rrp|$. Then, the Hartree-related term of the derivative discontinuity can be expressed as
\begin{equation}\label{eq.EH.phi.j}
\Delta_{H} = \frac{1}{2M^2} \sum\limits_{j} \int_{all} d^3r |\bar{\pphi}_j(\rr)|^2 V_{Hj}(\rr),
\end{equation}
where $j$ stands for $ho$ or $lu$ and $V_{Hj}(\rr) = \int_{all} d^3r' |\bar{\pphi}_j(\rrp)|^2 |\rr - \rrp|^{-1}$. In the limit of large $M$ (and neglecting the diverging term because the ``jellium'' background is irrelevant, as explained above), both $|\bar{\pphi}_j(\rr)|^2$ and $V_{Hj}(\rr)$ are periodic and remain finite as $M \rarr \infty$. Therefore the integration can be performed over a unit cell, yielding:  

\begin{equation}\label{eq.EH.phi.j.2}
\Delta_H = \frac{1}{2M} \sum\limits_{j} \int_{u.c.} d^3r |\bar{\pphi}_j(\rr)|^2 V_{Hj}(\rr).
\end{equation}
Therefore, the Hartree-related terms decay as $M^{-1}$ and vanish for the periodic solid.

The scaling of the xc contribution, $\Delta_{xc}$, is obviously much more interesting and it is here that the particular choice of the xc functional is crucial. For the exact xc functional, $\Delta_{xc}$ is generally expected to be non-vanishing, because $f_{xc}[n_0](\rr,\rrp) $ is known to exhibit divergence (see, e.g., the discussion in Ref.\ \cite{Onida2002,Botti2007}, and references therein). The nature of the singularity in the exact xc functional, therefore, must be such that for a periodic solid $\Delta_{xc}$ obtained from Eq.\ (\ref{eq.Delta.M}) is the exact one. Namely, the scaling for $\Delta_{xc}$ with $M$ as $M \rarr \infty$ should be $\sim M^0$. In parallel, the xc potential $v_{xc}$ must scale as $\sim M^0$, and the xc energy $E_{xc}$ as $\sim M^1$.

Unfortunately, this is not the case for simple functionals such as the LDA. In the LDA, the xc kernel can be expressed as $f_{xc}(\rr,\rrp) = g_{xc}(\rr) \delta(\rr - \rrp)$, where $g_{xc}(\rr)$ is a function of the density (and therefore periodic in a periodic system). As a result, the xc-related terms in Eq.~(\ref{eq.Delta.M}) simplify to $\frac{1}{2M} \int_{u.c.} d^3r |\bar{\pphi}_{j}(\rr)|^4 g_{xc}(\rr)$, i.e.\ they too decay as $M^{-1}$. Therefore, in the LDA approximation, for infinite systems $\Delta \sim M^{-1} \rarr 0$.

As a practical illustration of the above analysis, we used Eq.~(\ref{eq.Delta}) to evaluate $\Delta_H$ and $\Delta_{xc}$ in practice, focusing on GaAs as a prototypical example. Briefly, all electronic structure calculations were performed using the real-space PARSEC package~\cite{Chelikowsky1994,Chelikowsky1994a,Kronik2006,PARSEC_url}, while employing periodic boundary conditions \cite{Alemany2004,Natan2008a}. We used the Perdew-Zunger parameterization~\cite{PZ'81} of LDA with norm-conserving norm-conserving \cite{Hamann1979} Troullier-Martins pseudopotentials \footnote{The calculations were performed for GaAs in the zincblende crystal structure with the experimental lattice constant of $a = 10.684$ Bohr~\cite{Madelung2004}. A numerical precision of 0.02 eV in the reported energy gaps was obtained with a real-space grid spacing of $h = 0.25$ Bohr and an 11x11x11 $k$-point sampling scheme. The norm-conserving Troullier-Martins pseudopotentials [N.~Troullier and J.L.~Martins, Phys.\ Rev.\ B \textbf{43}, 1993 (1991)] for Ga and As were obtained using the APE program [M.J.~Oliveira and F.~Nogueira, Comput.\ Phys.\ Commun.\ \textbf{178}, 524 (2008), M.J.~Oliveira, F.~Nogueira, and T.~Cerqueira, APE -- Atomic Pseudopotential Engine, see http://www.tddft.org/programs/APE/], within the scalar-relativistic approximation, with the electronic configurations of [Ar]$4s^2 4p^1 4d^0$ and [Ar]$4s^2 4p^3 4d^0$ for Ga and As, respectively, with $s/p/d$ cutoff radii of 1.8/2.2/2.8 Bohr and 1.8/2.1/2.5 Bohr, using a non-linear core correction [S.G.~Louie, S.~Froyen, and M.~Cohen, Phys.\ Rev.\ B \textbf{26}, 1738 (1982)] and choosing the $d$ orbital as the local component in the Kleinman-Bylander projection scheme [L.~Kleinman and D.M.~Bylander, Phys.\ Rev.\ Lett.\ \textbf{48}, 1425 (1982)]. }. 

To investigate the dependence of $\Delta_H$, $\Delta_{xc}$, and $\Delta$ on the system size, we calculated these three quantities for increasingly large GaAs supercells of 1x1x1, 1x1x2, 1x1x3, 1x1x4, 1x2x2, and 2x2x2 primitive unit cells. Eq. (\ref{eq.Delta}) was then used under the assumption that $\pphi_{ho}$ and $\pphi_{lu}$ can be taken as the highest occupied and lowest unoccupied orbitals, respectively, of the supercell. 
\footnote{We draw special attention to the assumption underlying the derivation leading to Eq.~(\ref{eq.Delta}): the states with $N_0$, $N_0-1$, and $N_0+1$ electrons have to be pure, i.e.\ non-degenerate, states. Note that some degeneracies can be trivially removed: Degeneracy between energy levels in the two spin channels is removed by introducing an infinitesimal magnetic field; Degeneracy between symmetric k-points in a periodic crystal is removed by an infinitesimal spatial distortion of the crystal structure. However, non-trivial degeneracies, e.g.\ a simultaneous fractional occupation of $d$- and $s$-bands, requires a non-infinitesimal perturbation for their removal. For these cases, the ensemble approach should be generalized to include more than two components in Eqs.~(\ref{eq.Lambda}) and~(\ref{eq.LambdaKS}), which may affect the expression obtained for $\Delta$. Therefore, for cases of, e.g., metals and semi-metals a more general treatment is needed.}  
In other words, the supercell is treated as a {\it finite but topologically periodic} system and therefore approaches the bulk limit as the number of primitive unit cells, $m$, approaches infinity.

\begin{figure}
 \centering
  \includegraphics[scale=0.40,trim=35mm 0mm 0mm 0mm]{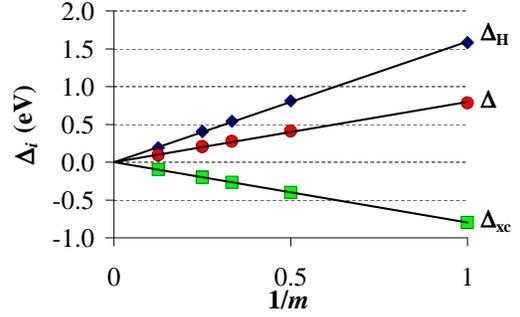}\\
  \caption{The derivative discontinuity $\Delta$, as well as its Hartree- and xc-components ($\Delta_H$ and $\Delta_{xc}$, respectively), as a function of $1/m$, where $m$ is the number of primitive unit cells in the supercell, for GaAs}\label{fig.Delta_i_GaAs}
\end{figure}

The results are given in detail in Fig.~\ref{fig.Delta_i_GaAs}. Clearly $\Delta_H$, $\Delta_{xc}$, and their sum, $\Delta$, are indeed all linear with $1/m$ and vanish in the limit of a large enough supercell. As the LDA-KS gap remains a constant $\sim 0.6$ eV for all $m$, in the bulk limit the corrected LDA gap simply approaches the uncorrected one. Similar trends were obtained for several other prototypical semiconductors (e.g., Si, Ge, InP) and are not shown for brevity. 

What is the physical origin for the apparent failure of the ensemble-correction for LDA (and indeed for any semi-local functional) in the limit of a periodic solid? To understand it, consider again the results of Sec.~\ref{sec.ensemble}, in particular Fig.~\ref{fig.piecewise.ensemble}. As shown there, the ensemble correction strongly reduces the curvature of the total energy versus particle number curve. This greatly assists in bringing the fundamental gap deduced from eigenvalue differences closer to the one deduced from total energy differences. That this would also help improve agreement with experiment hinges on the assumption that the fundamental gap deduced from total energy differences is close to the experimental one.
As discussed in Sec.~\ref{sec.ensemble}, for small and mid-size objects extensive numerical experience shows that this is often the case (see, e.g., \cite{Kaduk2012,Ogut97,Kronik2002a,Moseler2003,Weissker2004,Kr'10,
Argaman2013}). But for the infinite limit, it is in fact known that with the LDA, whose xc kernel is not singular, the fundamental gap deduced from total energy differences corresponds poorly to experiment and simply approaches the KS gap \cite{Godby1998} (For a similar reason, gaps deduced from time-dependent LDA also reduce to the LDA ones in the solid state limit -- see~\cite{Onida2002,Izmaylov2008}, and references therein). 

Therefore, the correction corresponding to LDA should indeed vanish. From that perspective one could argue that in the solid-state limit the ensemble correction scheme ``fails successfully'' for LDA, as it yields precisely what it was built to deliver -- consistency between total energy differences and eigenvalue differences (both of which are, alas, equally wrong).
Complementarily, several studies have shown that in the bulk limit the LDA total energy versus particle number curve is piecewise-linear even without ensemble corrections, albeit with the wrong slope \cite{Cohen12,Mori-Sanchez2008}. Also from this perspective, $\Delta$ must vanish, as there is no curvature to reduce. 

From yet a different perspective, mathematically the difficulty arises because the $ho$ and $lu$ orbitals are extremely delocalized, whereas the LDA xc kernel is extremely localized. This advocates the importance of ultra-non-local kernels. But in lieu of developing new functionals, another possibility is to localize the $ho$ and $lu$ orbitals. One such localization procedure is the above-mentioned dielectric screening  based one suggested by Chan and Ceder \cite{Chan2010} and others may be envisaged. In fact, one could argue that use of a small supercell as in Fig.~\ref{fig.Delta_i_GaAs} is, loosely speaking, a form of (uncontrolled) localization. Indeed if one were to take the results for the single unit cell literally, one would obtain $\Delta$=0.78 eV which would suggest a satisfying (but deceptive) agreement between the fundamental gap, $E_g^\KS + \Delta = 1.39$ eV and the experimental fundamental gap value, 1.51 eV~\cite{Madelung2004}. A similar behavior is obtained for other semiconductors as well. This suggests that controlled, physically justified localization procedures may prove to be key to systematic gap predictions even within LDA.

\section{Conclusions}

In this article, we have revisited the issue of the derivative discontinuity from an ensemble-DFT perspective. We have shown much of the deviation of approximate functionals from piecewise linearity is in fact due to the lack of an ensemble treatment. We have used this to show that \emph{all} exchange-correlation functionals possess a derivative discontinuity, which arises naturally from the application of ensemble considerations within DFT, without any empiricism or any further approximations beyond the choice of the xc functional. We then expressed this derivative discontinuity in closed form using only quantities obtained in the course of a standard DFT calculation of the neutral system. We showed that for small, finite systems, addition of this derivative discontinuity indeed results in a greatly improved prediction for the fundamental gap, even when based on the most simple approximate exchange-correlation density functional - the local density approximation (LDA). We then discussed the limit of an infinitely large system, so as to approach the solid-state limit. We found that the same scheme is exact in principle, but results in a vanishing derivative discontinuity correction when applied to semi-local functionals. This failure was shown to be directly related to the failure of semi-local functionals in predicting fundamental gaps from total energy differences in extended systems. Last, we discussed possible future remedies, especially usage of localization schemes.

\acknowledgments
We thank an anonymous Referee of Ref.~\cite{KrKronik13} for motivating the current work. We thank Patrick Rinke (Fritz-Haber-Institut, Berlin), Vojt\v{e}ch Vl\v{c}ek and Stephan K\"ummel (Bayreuth University), Richard M.\ Martin (University of  Illinois at Urbana-Champaign), Helen Eisenberg and Roi Baer (Hebrew University), and Ofer Sinai (Weizmann Institute) for helpful discussions. This work has been supported by the European Research Council and the Lise Meitner center for computational chemistry. E.K.\ is a recipient of the Levzion scholarship.

\clearpage

\bibliographystyle{apsrev_modified}
\bibliography{bibliography,library_SC}

\end{document}